\documentclass[twocolumn]{aastex61.mod}

\newcommand\aastex{AAS\TeX}

\received{December, 11 2017}
\revised{January, 2 2018}
\accepted{January, 3 2018}

\shorttitle{\aastex\ The \hsts large programme on $\omega$ Centauri -- III.}
\shortauthors{Libralato et al.}

\usepackage{xspace}
\usepackage{natbib}
\usepackage{amsmath}

\newcommand{\hstl}{\textit{Hubble Space Telescope}\xspace}
\newcommand{\hsts}{\textit{HST}\xspace}
\newcommand{\ocen}{$\omega$\,Cen\xspace}

\begin{document}

\title{The \hsts large programme on $\omega$ Centauri -- III. Absolute
  proper motion.}

\correspondingauthor{Mattia Libralato}
\email{libra@stsci.edu}

\author[0000-0001-9673-7397]{Mattia Libralato}
\affil{Space Telescope Science Institute 3700 San Martin Drive, Baltimore, MD 21218, USA}

\author[0000-0003-3858-637X]{Andrea Bellini}
\affil{Space Telescope Science Institute 3700 San Martin Drive, Baltimore, MD 21218, USA}

\author[0000-0003-4080-6466]{Luigi R. Bedin}
\affil{INAF-Osservatorio Astronomico di Padova, Vicolo dell'Osservatorio 5, Padova, I-35122, Italy}

\author{Edmundo Moreno D.}
\affil{Instituto de Astronom\'ia, Universidad Nacional Aut\'onoma de M\'exico, Apartado Postal 70-264, 04510-M\'exico DF, Mexico}

\author{Jos\'e G. Fern\'andez-Trincado}
\affil{Departamento de Astronom\'ia, Universidad de Concepci\'on,Casilla 160-C, Concepci\'on, Chile}
\affil{Institut Utinam, CNRS UMR6213, Univ. Bourgogne Franche-Comt\'e, OSU THETA, Observatoire de Besan\c{c}on, BP 1615, 25010 Besan\c{c}on Cedex, France}

\author[0000-0002-8195-8493]{Barbara Pichardo}
\affil{Instituto de Astronom\'ia, Universidad Nacional Aut\'onoma de M\'exico, Apartado Postal 70-264, 04510-M\'exico DF, Mexico}
  
\author[0000-0001-7827-7825]{Roeland P. van der Marel}
\affil{Space Telescope Science Institute 3700 San Martin Drive, Baltimore, MD 21218, USA}
\affil{Center for Astrophysical Sciences, Department of Physics \& Astronomy, Johns Hopkins University, Baltimore, MD 21218, USA}

\author[0000-0003-2861-3995]{Jay Anderson}
\affil{Space Telescope Science Institute 3700 San Martin Drive, Baltimore, MD 21218, USA}

\author[0000-0003-3714-5855]{D\'aniel Apai}
\affil{Department of Astronomy and Steward Observatory, The University of Arizona, 933 N. Cherry Avenue, Tucson, AZ 85721, USA}
\affil{Lunar and Planetary Laboratory, The University of Arizona, 1640 E. University Blvd., Tucson, AZ 85721, USA}

\author[0000-0002-6523-9536]{Adam J. Burgasser}
\affil{Center for Astrophysics and Space Science, University of California San Diego, La Jolla, CA 92093, USA}

\author{Anna Fabiola Marino}
\affil{Research School of Astronomy \& Astrophysics, Australian National University, Canberra, ACT 2611, Australia}

\author[0000-0001-7506-930X]{Antonino P. Milone}
\affil{Dipartimento di Fisica e Astronomia ``Galileo Galilei'' - Universit\`a di Padova, Vicolo dell'Osservatorio 3, Padova, IT-35122, Italy}

\author{Jon M. Rees}
\affil{Department of Astronomy and Steward Observatory, The University of Arizona, 933 N. Cherry Avenue, Tucson, AZ 85721, USA}

\author[0000-0002-1343-134X]{Laura L. Watkins}
\affil{Space Telescope Science Institute 3700 San Martin Drive, Baltimore, MD 21218, USA}



\begin{abstract}

In this paper we report a new estimate of the absolute proper motion
(PM) of the globular cluster NGC~5139 (\ocen) as part of the \hsts
large program GO-14118+14662. We analyzed a field 17 arcmin South-West
of the center of \ocen and computed PMs with an epoch span of
$\sim$15.1 years. We employed 45 background galaxies to link our
relative PMs to an absolute reference-frame system. The absolute PM of
the cluster in our field is:\ $(\mu_\alpha \cos\delta , \mu_\delta) =
(-3.341 \pm 0.028 , -6.557 \pm 0.043)$ mas yr$^{-1}$. Upon correction
for the effects of viewing perspective and the known cluster rotation,
this implies that for the cluster center of mass $(\mu_\alpha
\cos\delta , \mu_\delta) = (-3.238 \pm 0.028, -6.716 \pm 0.043)$ mas
yr$^{-1}$. This measurement is direct and independent, has the highest
random and systematic accuracy to date, and will provide an external
verification for the upcoming Gaia Data Release 2. It also differs
from most reported PMs for \ocen in the literature by more than
5$\sigma$, but consistency checks compared to other recent catalogs
yield excellent agreement. We computed the corresponding
Galactocentric velocity, calculated the implied orbit of \ocen in two
different Galactic potentials, and compared these orbits to the orbits
implied by one of the PM measurements available in the literature. We
find a larger (by about 500 pc) perigalactic distance for \ocen with
our new PM measurement, suggesting a larger survival expectancy for
the cluster in the Galaxy.

\end{abstract}

\keywords{globular clusters: individual (NGC~5139 (\ocen)) --
  astrometry -- proper motions -- Galaxy: kinematics and dynamics }



\defcitealias{1965RGOB..100...81M}{M65}
\defcitealias{1966ROAn....2....1W}{W66}
\defcitealias{1999AJ....117..277D}{D99}
\defcitealias{2000AA...360..472V}{vL00}
\defcitealias{2002ASPC..265..399G}{G02}
\defcitealias{2006AA...445..513V}{vdV06}
\defcitealias{2013AA...558A..53K}{K13}
\defcitealias{2017MNRAS.469..800M}{Paper~I}
\defcitealias{2018arXiv180101504B}{Paper~II}

\section{Introduction}

NGC~5139 (\ocen) is one of the most complex and intriguing globular
clusters (GCs) in our Galaxy. It was the first GC known to contain
multiple stellar populations
\citep{1997phd..jay....ander,2004ApJ...605L.125B}, and its populations
show increasing complexity the more we study them \citep[e.g.,][and
  references
  therein]{2017ApJ...842....6B,2017ApJ...842....7B,2017ApJ...844..164B,2017MNRAS.464.3636M}. In
this context, the \hsts large program GO-14118+14662 (PI: Bedin),
aimed at studying the white-dwarf cooling sequences of \ocen
\citep[see][]{2013ApJ...769L..32B}, began observations in 2015.

In Paper~I of this series \citep{2017MNRAS.469..800M}, we traced and
chemically tagged for the first time the faint main sequences of \ocen
down to the Hydrogen-burning limit. In Paper~II
\citep{2018arXiv180101504B} we began to investigate the internal
kinematics of the multiple stellar population of the cluster.

In this third study of the series, we present an analysis of the
absolute proper motion (PM) of \ocen. Despite long being considered a
GC, \ocen has also been thought to be the remnant of a
tidally-disrupted satellite dwarf galaxy
\citep[e.g.,][]{2003MNRAS.346L..11B}. As such, understanding the orbit
of this object is an important step toward understanding its true
nature.

One of the key pieces of information needed to trace the orbit of an
object is its absolute motion in the plane of the sky, i.e., its
absolute PM. Most of the current estimates of the PM of \ocen were
obtained with ground-based photographic plates \citep[e.g.][hereafter
  D99 and vL00, respectively]{1999AJ....117..277D,2000AA...360..472V},
which have very different systematic issues than space-based
measurements.

Thanks to the large number of well-measured background galaxies in our
deep \hsts images (Sect.~\ref{red}), we are able to determine a new,
independent measurement of the absolute PM of \ocen
(Sect.~\ref{apms}). The PM value we computed differs by more than
5$\sigma$ from the most recent estimates (Sect.~\ref{lit}). Our new
measurement also implies a different orbit for the cluster. As such,
we also computed orbits for \ocen with different Galactic potentials
and compared them with those obtained by employing a previous PM
estimate available in the literature (Sect.~\ref{orbit}).

\section{Data sets and reduction.}\label{red}

This study is focused on a $2.73 \times 2.73$ arcmin$^2$ field
\citepalias[named ``F1'' in][]{2018arXiv180101504B} at about 17 arcmin
South-West of the center of \ocen\footnote{$(\alpha,\delta)_{\rm
    J2000}$ $=$ ($13^{\rm h}26^{\rm m}47^{\rm
    s}\!\!.24$,$-47^\circ28^\prime46^{\prime\prime}\!\!.45$),
  \citet{2010ApJ...710.1032A}.}. The center of the field F1 is located
at $(\alpha,\delta)_{\rm ICRS}$ $\sim$ ($13^{\rm h}25^{\rm m}37^{\rm
  s}\!\!.32$,$-47^\circ40^\prime00^{\prime\prime}\!\!.21$). The field
was observed with \hsts's Ultraviolet-VISible (UVIS) channel of the
Wide-Field Camera 3 (WFC3) between 2015 and 2017 during GO-14118+14662
(PI: Bedin). Additional archival images were taken with the Wide-Field
Channel (WFC) of the Advanced Camera for Surveys (ACS) in 2002
(GO-9444, PI: King) and in 2005 (GO-10101, PI: King). As such, the
maximum temporal baseline covered by our observations is about 15.1
yr. The list of the data sets\footnote{DOI reference:
  \protect\dataset[10.17909/T9FD49]{http://dx.doi.org/10.17909/T9FD49}.}
employed here is summarized in Table~1 of
\citetalias{2018arXiv180101504B}. Note that, as part of the large
program GO-14118+14662, there are also WFC3 exposures obtained with
the near-infrared (IR) channel. However, we did not use near-IR data
because of the worse astrometric precision of WFC3/IR channel as
compared to that of ACS/WFC and WFC3/UVIS detectors.

\begin{figure*}
  \centering
  \includegraphics[trim=20 160 30 360,clip=true,scale=0.91,keepaspectratio]{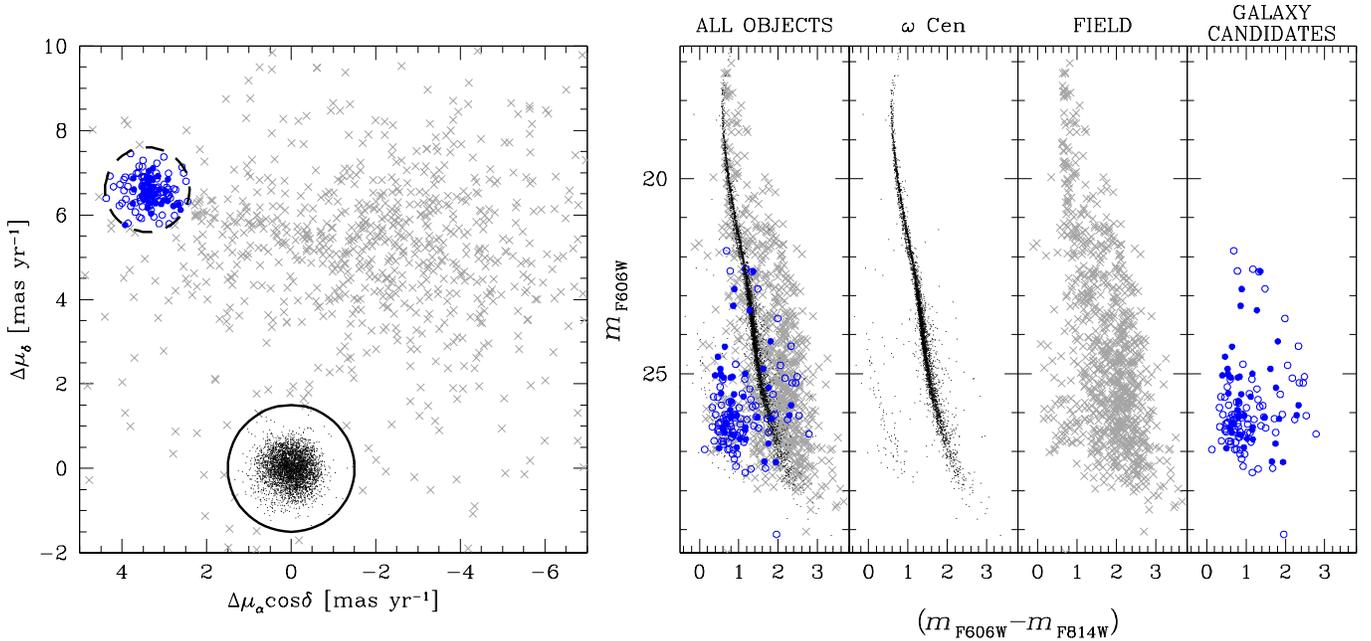}
  \caption{\textit{Left}: VPD of the relative PMs of the analyzed
    field. Black dots (within a circle of 1.5 mas yr$^{-1}$ from the
    origin of the VPD) are members of \ocen. Field stars are plotted
    with gray crosses. Objects classified as galaxies in our study are
    shown as blue circles. The black, dashed circle enclosing these
    galaxies is centered at $(3.4,6.6)$ mas yr$^{-1}$ and has a radius
    of 1 mas yr$^{-1}$. Solid (open) symbols are galaxies employed
    (discarded) in our estimate of the absolute PM of \ocen (see
    Sect.~\ref{apms} for detail). \textit{Right}: Color-magnitude
    diagrams of all objects in the analyzed field.}\label{fig1}
\end{figure*}

A detailed description of the data reduction and relative-PM
computation can be found in \citetalias{2018arXiv180101504B}. In the
following, we provide a brief overview.

We performed our analysis on the \texttt{\_flt} exposures corrected by
the official pipeline for charge-transfer-efficiency (CTE) defects
(see \citealt{2010PASP..122.1035A} for the description of this
empirical correction). For each camera/filter combination, we
extracted positions and fluxes of all detectable sources by
point-spread-function (PSF) fitting in a single finding wave without
removing the neighbor stars prior to the fit.  We used empirical,
spatial- and time-varying PSFs obtained by perturbing the publicly
available\footnote{\href{http://www.stsci.edu/~jayander/STDPSFs/}{http://www.stsci.edu/$\sim$jayander/STDPSFs/}
  .} \hsts ``library'' PSFs
\citep[see][]{2017ApJ...842....6B}. Finally, we corrected the stellar
positions for geometric distortion using the solutions of
\citet{2006acs..rept....1A}, \citet{2009PASP..121.1419B} and
\citet{2011PASP..123..622B}.

We set up a common reference-frame system (``master frame'') for each
epoch/camera/filter, which is used to cross-correlate the available
single-exposure catalogs. We made use of the Gaia Data Release 1
\citep[DR1,][]{2016AA...595A...1G,2016AA...595A...2G} catalog to (i)
set a master-frame pixel scale of 40 mas pixel$^{-1}$ (that of
WFC3/UVIS detector) and (ii) orient our system with X/Y axes parallel
to East/North. Then, we iteratively cross-identified the stars in each
catalog with those on the master frame by means of six-parameter
linear transformations.

Following \citet{2017ApJ...842....6B}, we performed a ``second-pass''
photometry stage by using the software \texttt{KS2}, an evolution of
the code used to reduce the ACS Globular Cluster Treasury Survey data
in \citet{2008AJ....135.2055A}, and we measured stellar positions and
fluxes simultaneously by using individual epochs/filters/exposures at
once. We set \texttt{KS2} to examine F606W- and F814W-filter images to
detect every possible object, and then to measure them also in all
other filters. Stellar positions and fluxes were measured by
re-fitting the PSF in the neighbor-subtracted images. Differently from
\citet{2017ApJ...842....6B}, we are interested not only in the stellar
members of \ocen, but also field objects (stars and galaxies) that
might have moved with respect to \ocen by more than 2 pixels (the
\texttt{KS2} method \#1 searching radius) between two
epochs. Therefore, we executed \texttt{KS2} independently for the
GO-9444+10101 and GO-14118+14662 data to avoid mismatching
objects. Finally, our instrumental magnitudes were calibrated onto the
Vega magnitude system.

The \texttt{KS2} software also provides positions and fluxes for
detected objects in each raw exposure. Using neighbor-source
subtraction, crowded-field issues that might affect the first-pass
approach were reduced. Therefore, we employed these raw,
\texttt{KS2}-based catalogs to measure the PMs instead of those
produced by the first-pass photometry.

PMs were computed using the same methodology described in
\citet{2014ApJ...797..115B}. The cornerstones of their iterative
procedure are (i) transforming the position of each object as measured
in the single catalog/epoch on to the master-frame system, and (ii)
fitting a straight line to these transformed positions as a function
of the epoch. The slope of this line, computed after several
outlier-rejection stages, is a direct measurement of the PM. We refer
to \citet{2014ApJ...797..115B} and \citetalias{2018arXiv180101504B}
for a detailed description of the PM extraction and systematic
corrections.

Because the transformations that relate positions measured in one
exposure to those in another were based on cluster stars, by
construction, the PM of each star is measured with respect to the
average cluster motion. Therefore, the distribution of \ocen stars is
centered on the origin of the vector-point diagram (VPD), while
foreground/background objects are located in different parts of the
VPD.

The relative PMs\footnote{To avoid confusion between relative and
  absolute PMs, we use the notation ($\Delta \mu_\alpha \cos\delta$,
  $\Delta \mu_\delta$) for the former and ($\mu_\alpha \cos\delta$,
  $\mu_\delta$) for the latter.} of the objects in our field are shown
in Fig.~\ref{fig1}. Three groups of objects are clearly
distinguishable in the VPD: ~\\
1) cluster stars, centered in the origin of the VPD, with observed
dispersion of about 0.34 mas yr$^{-1}$ along both $\Delta \mu_\alpha
\cos\delta$ and $\Delta \mu_\delta$ (black points);~\\
2) background galaxies, clustered in a well-defined location $\sim
7.5$ mas yr$^{-1}$ from the mean motion of \ocen with a dispersion of
0.40 and 0.32 mas yr$^{-1}$ along $\Delta \mu_\alpha \cos\delta$ and
$\Delta \mu_\delta$, respectively (blue circles);~\\
3) field stars, characterized by a broad distribution at positive
$\Delta \mu_\delta$ and large negative $\Delta \mu_\alpha \cos\delta$,
and a narrow tail ending at the galaxy location (gray crosses).\\

In \citetalias{2018arXiv180101504B} we focused on the internal
kinematics of the multiple stellar populations hosted in \ocen, so
that only cluster stars with high-precision PMs were analyzed. In this
paper we focus on the global, absolute motion of the
cluster. Therefore, we require high accuracy but not necessarily high
precision, as was the case in \citetalias{2018arXiv180101504B}. As a
consequence, we make use of all objects with a PM measurement in our
catalog.

\begin{figure*}
  \centering
  \includegraphics[trim=30 170 30 350,clip=true,scale=0.92,keepaspectratio]{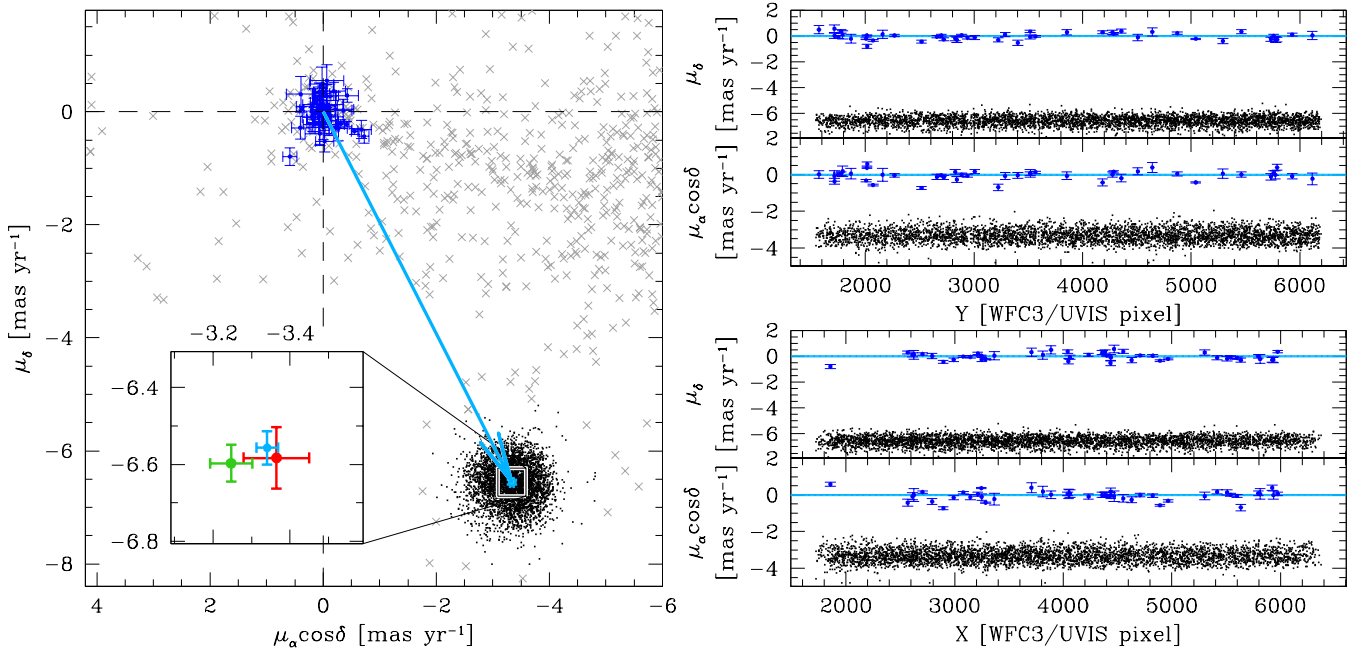}
  \caption{\textit{Left}: VPD of the absolute PMs in equatorial
    coordinates of \ocen. Black dots and gray crosses represent \ocen
    stars and field objects, respectively. The blue points (with error
    bars) are the galaxies actually used to compute the absolute
    PM. The azure arrow indicates the motion of \ocen stars with
    respect to the (fixed) galaxies, the error bars our $\sigma$
    uncertainty in the absolute PM value. In the inset, the green and
    red points (with error bars) are the absolute PMs measured using
    only F606W- and F814W-filter exposures, respectively (see text for
    detail). \textit{Right}: $\mu_\alpha cos\delta$ and $\mu_\delta$
    as a function of X (bottom panels) and Y (top panels) master-frame
    positions. We omitted field stars for clarity. The dashed, azure
    stripes represent the absolute PM $\pm \sigma$ error in each
    component. No systematic trend arises from these plots, proof of
    the robustness of our PMs.}\label{fig2}
\end{figure*}

\section{Absolute proper motions}\label{apms}

The first step is to link our \textit{relative} PMs to an
\textit{absolute} reference-frame system. This can be done in two
ways: indirectly (by relying on an external catalog of absolute
positions), and directly (by computing PMs with respect to fixed
sources in the sky, e.g., galaxies).

For the former approach, we can cross-identify cluster members with
catalogs providing absolute PMs like the \textit{Hipparcos}-Tycho
\citep{2000AA...355L..27H} or the UCAC5 \citep{2017AJ....153..166Z},
and compute the zero-point offset to be applied to our
PMs. Unfortunately, the only stars in common with the UCAC5 catalog
are saturated even in the short exposures and are not suitable for
astrometry.

Alternatively, we could use the Gaia catalog as a reference. Although
intriguing, the major drawback is that only 130 stars are found in
common between our catalog and Gaia, and none of them is present in
the Tycho-Gaia Astrometric Solution (TGAS), thus preventing us to
indirectly register our PMs into an absolute reference frame. Note
that TGAS sources would be saturated in our images as is the case of
the objects in common with UCAC5 catalog, hence not useful for the
registration.

The direct way consists in estimating the absolute PM from our data
set using background galaxies in the field as a reference (cf.,
\citealt{2003AJ....126..247B,2006AA...460L..27B};
\citealt{2006AA...456..517M}; \citealt{2010AA...513A..51B};
\citealt{2013ApJ...779...81M}; \citealt{2015ApJ...803...56S}), which
is the method we followed.

We started by identifying by eye the center of the distribution of the
many galaxy-like objects in the VPD (see Fig.~\ref{fig1}). We defined
a sample as all objects within a circle of 1 mas yr$^{-1}$ from the
preliminary center of that group (black, dashed circle in
Fig.~\ref{fig1}). We then visually inspected the stacked images and
rejected all fuzzy galaxies (i.e., objects without a well-defined,
point-source-like core), object close to saturated stars or their
bleeding columns, and sources which could not be excluded as stars. At
the end of our purging process, we are left with 45 galaxies, whose
trichromatic rasters are shown in appendix \ref{A1}. Given the
tightness of the galaxy distribution in the VPD, it is clear that the
positions of these galaxies are measured quite well, even compared to
the more plentiful cluster members.

We measured the 5$\sigma$-clipped median value of the PMs in each
coordinate for the surviving galaxies. This median value (with the
opposite sign) represents the absolute PM of \ocen stars at the
position of our field F1 (Fig.~\ref{fig2}). We find:
\begin{equation*}
    (\mu_\alpha \cos\delta , \mu_\delta)_{\rm @F1}
\end{equation*}
\vskip -15pt
\begin{equation}
  =
\end{equation}
\vskip -20pt
\begin{equation*}
    (-3.341\ \pm\ 0.028 , -6.557\ \pm\ 0.043) \textrm{ mas  yr$^{-1}$ .}
\end{equation*}
The quoted errors are the standard error to the mean.

In the right-hand panels of Fig.~\ref{fig2} we also show the absolute
PMs along the $\mu_\alpha \cos\delta$ and $\mu_\delta$ directions as a
function of the X and Y positions on the master frame. We find no
systematic trends in the PM distribution of either galaxies or \ocen
stars. Magnitude- and color-dependent systematic errors are also not
present in our PMs as shown in Fig.~3 of
\citetalias{2018arXiv180101504B}.

The cores of our selected galaxies are resolved by \hsts, and the
galaxies themselves may have different flux distributions and shapes
if observed through different filters. As such, the centroid of a
galaxy could differ if measured with different filters, resulting in
biased PM measurements. Furthermore, \citet{2011PASP..123..622B} noted
that blue and red photons are refracted differently when they pass
through the fused-silica window of the CCD of the WFC3/UVIS detector,
even if such effect is marginal for filters redder than 3000 \AA.

To assess the contributions of these effects to our absolute-PM
estimate, we re-computed the relative PMs as described in
\citetalias{2018arXiv180101504B} in two ways: (1) using only F606W
images, and (2) using only F814W images.  We then measured the
absolute PM of the cluster in both cases, and compared the results
with our original estimate. In the inset of the left-hand panel of
Fig.~\ref{fig2}, we plot the original (in azure), F606W- (in green)
and F814W-based (in red) absolute PMs for the cluster. The three
absolute PMs are in agreement (well within the error bars), meaning
that the galaxy morphology and color effects have negligible impacts
in our measurements.

Our observations were obtained in different periods of the year and,
as such, our absolute-PM estimate might include an annual-parallax
term. However, the contribution of the annual parallax would be of
about 0.025 mas yr$^{-1}$ at most \citep[by assuming a distance of
  \ocen of 5.2 kpc,][2010 edition]{1996AJ....112.1487H}, a value of
the order of our absolute-PM uncertainties.

Furthermore, it is worth noting that our observations were achieved in
July (2002), December (2005) and August (2015 and 2017). The highest
contribution of the annual parallax would take place if the PM of the
galaxies (the parallax effect on \ocen stars reflects on background
objects because we computed the PMs in a relative fashion) was
obtained by employing 2002--2005 or 2005--2015 (or 2017) epoch
pairs. However, the PM of 43 out of the 45 galaxies we used in our
absolute-PM analysis was computed by adopting all observations at our
disposal, for a total temporal baseline of 15.1 yr. Therefore, the fit
of the PM is well constrained by the first- and last-epoch
observations (at about the same phase in the Earth orbit), and the
slope of the straight line (the PM) is expected to be affected by a
term much lower than 0.025 mas yr$^{-1}$. For all these reasons, we
expect the annual parallax to be negligible and we chose not to
correct it.

\subsection{Cluster rotation}\label{rot}

It is well-established that \ocen rotates in the plane of the sky
(e.g., \citetalias{2000AA...360..472V}; \citealt{2001ASPC..228...43F};
\citealt{2002ASPC..265...41V}; \citealt{2006AA...445..513V}, hereafter
vdV06). We can also infer the presence of rotation directly from our
VPDs. \citet{2017ApJ...850..186H} showed that the skeweness in the PMs
of 47\,Tuc is a signature of the differential rotation of the cluster,
confirming previous results of the multiple-field PM analysis made by
\citet{2017ApJ...844..167B}. As discussed in
\citetalias{2018arXiv180101504B}, the PMs of \ocen are also skewed in
the tangential direction, consistent with cluster rotation.

Therefore, the PM of \ocen we obtained in our analysis is a
combination of the center of mass (COM) motion and the internal
cluster rotation at the position of the field F1. The four fields
observed in our \hsts large program of \ocen are located at different
position angles and distances from the cluster center (see Fig.~1 of
\citetalias{2018arXiv180101504B}), and observations for the other
three fields are still ongoing. When all the scheduled multiple-epoch
observations are completed for all the fields, we will be able to make
a precise measurement of the rotation of the cluster. Here, we use the
kinematic analyses of \citetalias{2006AA...445..513V}, in which they
adopted the PM catalog of \citetalias{2000AA...360..472V} to derive
the rotation of \ocen.

Our field is 17 arcmin away from the center of \ocen, and projection
effects (because of the different line of sight between the COM and
our field F1) can produce an additional \textit{perspective} PM
component. In order to measure the PM of the COM of \ocen, we must
also remove the contribution of the projection from our estimate.

The size of our field is small as compared to its distance from the
center of \ocen. Therefore, we assumed that, at the first order, the
effects of perspective and internal rotation are the same for all
stars in field F1. Because of this, we corrected the absolute-PM
measurement we inferred in Sect.~\ref{apms} and not the PM of each
individual star.

First, we computed the contribution of perspective using the
prescription given in \citet{2002AJ....124.2639V}. By adopting as the
reference position of our field the average position of all its stars,
we find:
\begin{equation*}
    (\mu_\alpha \cos\delta , \mu_\delta)_{\rm Perspective}
\end{equation*}
\vskip -15pt
\begin{equation}
  =
\end{equation}
\begin{equation*}
  (0.008 , 0.043) \textrm{ mas  yr$^{-1}$ .}
\end{equation*}
We then assessed the contribution of the cluster rotation. Table~3 of
\citetalias{2006AA...445..513V} reports the mean PM of \ocen in
different polar sectors in the plane of the sky. These velocities size
the amount of the rotation in different locations of the
cluster. After we identified in which polar sector our field lies, we
transformed the corresponding PM values in Table~3 of
\citetalias{2006AA...445..513V} from the axisymmetric reference frame
adopted in their study to the equatorial reference frame, and
decomposed these PMs in the corresponding radial and tangential
components. Since we are only interested in the amount of rotation, we
just need to correct our PMs by the tangential component. Finally, we
transformed these PM values back to the equatorial coordinate
system. The expected contribution of cluster rotation to the absolute
PM is:
\begin{equation*}
    (\mu_\alpha \cos\delta , \mu_\delta)_{\rm Rot}
\end{equation*}
\vskip -15pt
\begin{equation}
  =
\end{equation}
\begin{equation*}
  (-0.111 , 0.116) \textrm{ mas  yr$^{-1}$ .}
\end{equation*}
By subtracting the perspective and the rotation components to our
absolute PM at the position of our field F1, we obtain our best
estimate of the absolute PM of the COM of \ocen:
\begin{equation*}
    (\mu_\alpha \cos\delta , \mu_\delta)_{\rm COM}
\end{equation*}
\vskip -15pt
\begin{equation}
  =
\end{equation}
\begin{equation*}
  (-3.238 \pm 0.028 , -6.716 \pm 0.043) \textrm{ mas  yr$^{-1}$ .}
\end{equation*}
The difference between the absolute PM at the position of our field F1
and that of the COM is of about 0.19 mas yr$^{-1}$, dominated by
cluster rotation.  This correction is qualitatively in agreement with
that expected by Fig.~6 of \citetalias{2006AA...445..513V}, both in
magnitude and direction.

It is important to note that \citetalias{2006AA...445..513V} employed
the PM catalog of \citetalias{2000AA...360..472V}, which is known to
have severe color- and magnitude-dependent systematic effects as
described by \citet{2003ApJ...591L.127P}. The uncertainty in the COM
PM listed above includes only random errors, and no systematic
uncertainties in the rotation correction. If the $\sim 4$ km s$^{-1}$
PM rotation measured by \citetalias{2006AA...445..513V} is inaccurate
by more than $\sim 30$\%, then this systematic error dominates the
error budget.

To determine the corresponding Galactocentric velocity of the COM, we
adopt a Cartesian Galactocentric coordinate system ($X, Y, Z$), with
the origin at the Galactic Center, the $X$-axis pointing in the
direction from the Sun to the Galactic Center, the $Y$-axis pointing
in the direction of the Sun's Galactic rotation, and the $Z$-axis
pointing towards the Galactic North Pole. We use the same quantities
and corresponding uncertainties for the Sun and for \ocen as in
Sect.~\ref{orbit} below (see Tables~\ref{tab2} and \ref{tabkin}). This
implies a Galactocentric $(X, Y, Z)$ position
\begin{equation}
  {\vec r} = (-5.1, -3.9, 1.3) \textrm{ kpc ,}
\end{equation}
and a Galactocentric velocity vector
\begin{equation}
  {\vec v} = (97.9 \pm 3.0, -18.0 \pm 8.6, -80.2 \pm 7.2) \textrm{ km s$^{-1}$ .}
\end{equation}
The corresponding Galactocentric radial and tangential velocities are
\begin{equation}
  (V_{\rm rad}, V_{\rm tan}) =
   (-82.0 \pm 7.8, 98.1 \pm 10.4) \textrm{ km s$^{-1}$ ,}
\end{equation}
and the observed total velocity with respect to the Milky Way is
\begin{equation}
  v \equiv |{\vec v}| = (127.9 \pm 3.5) \textrm{ km s$^{-1}$ .}
\end{equation}
The listed uncertainties were obtained from a Monte-Carlo scheme that
propagates all observational distance and velocity uncertainties and
their correlations, including those for the Sun.

\subsection{Field stars: who is who}\label{field}

\begin{figure}
  \centering
  \includegraphics[trim=25 145 125 90,clip=true,width=\columnwidth]{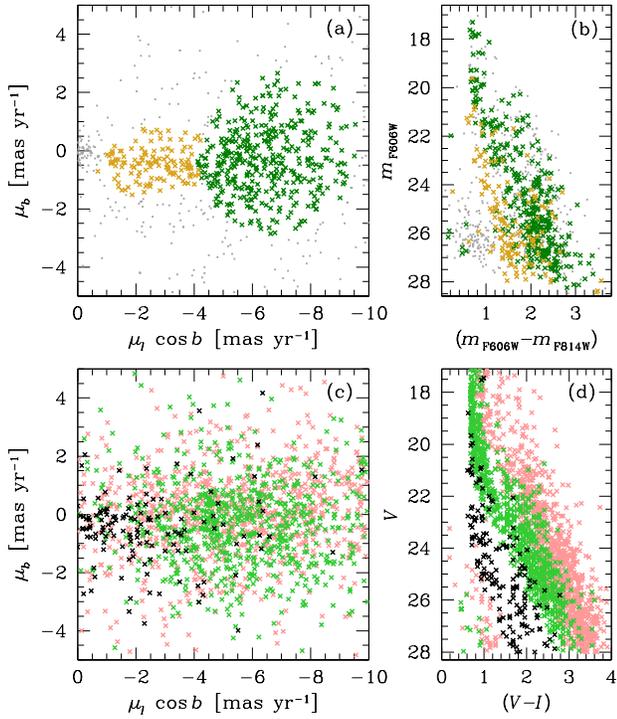}
  \caption{Comparison between observed (top panels) and simulated
    (bottom panels) VPDs in Galactic coordinates (left-hand panels)
    and color-magnitude diagrams (right-hand panels). In panels (a)
    and (b) we show galaxies/field stars with gray dots and highlight
    the two investigated groups in gold and dark green. In panels (c)
    and (d) we color-code the simulated halo, thick and thin disk
    stars in black, light green and pink, respectively.}\label{fig4}
\end{figure}

\begin{figure}
  \centering
  \includegraphics[trim=10 140 40 120,clip=true,width=1.02\columnwidth]{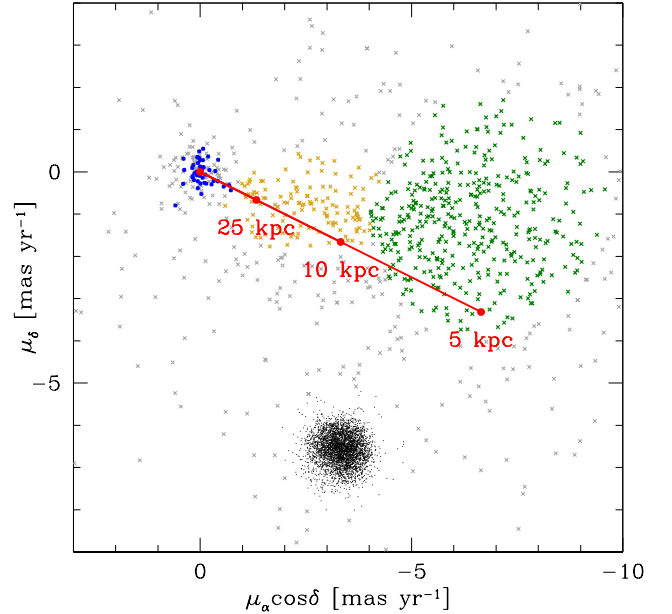}
  \caption{VPD of the absolute PMs in equatorial coordinates
    highlighting the galaxies (in blue) and the two groups of field
    stars analyzed in Sect.~\ref{field} (in gold and dark-green
    colors). \ocen stars and the remaining field objects are shown
    with black dots and gray crosses, respectively. The red line marks
    the expected reflex motion of stars at different Galactocentric
    distances (see text for detail).}\label{fig5}
\end{figure}

The VPDs in Figs.~\ref{fig1} and \ref{fig2} reveal the presence of two
main field-star components. In panel (a) of Fig.~\ref{fig4} we
highlighted them with gold and dark-green crosses. These two groups of
stars also show different loci in the color-magnitude diagrams (panel
b). To identify the stellar populations they belong to, we compared
the observed absolute PMs of field stars in our catalog to those of
the Besan\c{c}on models \citep{2003AA...409..523R}. We simulated stars
out to 50 kpc in a $6 \times 6$ arcmin$^2$ region centered on our
field. We assumed the photometric and PM errors to be zero since we
are only interested in tagging each component of the field. The final
simulated sample is composed of halo (shown with black crosses in the
bottom panels of Fig.~\ref{fig4}), thin- (pink crosses), and
thick-disk (light-green crosses) stars.

Both color-magnitude diagrams and VPDs (in Galactic coordinates) in
Fig.~\ref{fig4} suggest that the broad group of field stars in our
observed VPD is mainly constituted by disk stars, while the narrow
stream connecting the distributions of disk stars to the galaxies is
associated with halo stars.

En passant, we find broad agreement between observed and simulated
PMs, an independent validation of the theoretical predictions of
Besan\c{c}on models.

We can reach similar conclusions by considering that, when we look at
the stars along the line of sight toward \ocen, we see reflex motion
due to the motion of the Sun. By assuming a Sun motion in
Galactocentric reference frame given in Sect.~\ref{orbit} below (see
Table~\ref{tab2}), we expect a reflex proper motion of:
\begin{equation}
  \left\{
  \begin{array}{rl}
  \mu_\alpha \cos\delta & = -6.391 \times (5.2/d) \textrm{ mas  yr$^{-1}$} \\
  \mu_\delta & = -3.190 \times (5.2/d) \textrm{ mas  yr$^{-1}$ ,}
  \end{array}
  \right.
\end{equation}
with $d$ the distance of a given star from the Galactic center in
kpc. For hypothetical stars that stand still in the Galactocentric
reference frame, this motion is equal to the observed PM. In the VPD
in Fig.~\ref{fig5}, we used the equations above to mark the expected
location of a given object at different Galactocentric distances.

Galaxies and very-distant halo stars have no motions in the
Galactocentric reference frame. Therefore, they fall at the large
distance end of the red line with a small scatter (because the PMs
induced by their peculiar velocities at large distances are
small). Intermediate-distance stars are in the inner disk and bulge,
and these stars have a orbital motion in the Milky Way. As such, they
do not fall exactly along the line. Finally, nearby stars have a large
scatter about the line because of their large PMs induced by peculiar
velocities, even if the latter are relatively small.

Looking at Fig.~\ref{fig5}, it seems that gold points are much closer
to the straight line and at large distances, as expected by halo and
intermediate-disk stars. Green objects have instead a broad
distribution far from the red line, suggesting that they are nearby
disk stars.

\begin{table*}
  \begin{center}
    \caption{List of the absolute PM of \ocen in the literature and computed in this paper.} \label{tab1}
    \begin{tabular}{clrrcc}
      \tableline
      \tableline
      \textbf{ID} & \textbf{Reference} & \textbf{$\bf \mu_\alpha \cos \delta$} & \textbf{$\bf \mu_\delta$} & \textbf{Extragalactic?} & \textbf{Notes} \\
      & & \textbf{[mas yr$^{-1}$]} & \textbf{[mas yr$^{-1}$]} & & \\
      \tableline
      \tableline
      \citetalias{1965RGOB..100...81M} & \citet{1965RGOB..100...81M} & $+0.5 \pm 0.6$ & $-7.7 \pm 0.5$ & No & \\
      \citetalias{1966ROAn....2....1W} & \citet{1966ROAn....2....1W} & $-3.6 \pm 0.5$ & $-6.0 \pm 0.5$ & No & (\textit{a}) \\
      \citetalias{1999AJ....117..277D} & \citet{1999AJ....117..277D} & $-4.88 \pm 0.35$ & $-3.47 \pm 0.34$ & Yes & (\textit{b}) \\
                                       &                             & $-5.08 \pm 0.35$ & $-3.57 \pm 0.34$ & Yes & \\
      \citetalias{2000AA...360..472V}  & \citet{2000AA...360..472V}  & $-3.97 \pm 0.41$ & $-4.38 \pm 0.41$ & No & (\textit{c}) \\
      \citetalias{2002ASPC..265..399G} & \citet{2002ASPC..265..399G} & $-4.2 \pm 0.5$ & $-5.1 \pm 0.5$ & No & \\
                                       &                             & $-4.1 \pm 0.5$ & $-4.8 \pm 0.5$ & No & (\textit{d}) \\
      \citetalias{2013AA...558A..53K}  & \citet{2013AA...558A..53K}  & $-6.01 \pm 0.25$ & $-5.02 \pm 0.25$ & No & (\textit{e}) \\
      \tableline
      vL00 (@F1) & This paper & $-4.17 \pm 0.41$ & $-4.31 \pm 0.41$ & No & (\textit{f}) \\
      vL00 (COM) & This paper & $-4.06 \pm 0.41$ & $-4.43 \pm 0.41$ & No & (\textit{g}) \\
      \tableline
      HSOY (@F1) & This paper & $-4.08 \pm 0.44$ & $-6.68 \pm 0.26$ & No & \\
      HSOY (All) & This paper & $-3.21 \pm 0.15$ & $-6.51 \pm 0.14$ & No & (\textit{h})\\
      UCAC5 (@F1) & This paper & $-4.10 \pm 0.27$ & $-7.00 \pm 0.26$ & No & \\
      UCAC5 (All) & This paper & $-3.30 \pm 0.03$ & $-6.80 \pm 0.03$ & No & (\textit{h}) \\
      \tableline
      L18 (@F1) & This paper & $-3.341 \pm 0.028$ & $-6.557 \pm 0.043$ & Yes \\
      L18 (COM)  & This paper & $-3.238 \pm 0.028$ & $-6.716 \pm 0.043$ & Yes & (\textit{i}) \\
      \tableline
      \tableline
    \end{tabular}
  \end{center}
  \tablecomments{(\textit{a}){: Obtained by \citet{2002ASPC..265..399G} by linking \citet{1965RGOB..100...81M} and \citet{1966ROAn....2....1W} catalog on Tycho 2 catalog.} (\textit{b}){: The two values corresponds to the absolute PM with and without the correction for the field gradient of the mean motion of the reference stars, respectively.} (\textit{c}){: Absolute-PM registration on Tycho 2 catalog.} (\textit{d}){: Derived using UCAC1 \citep{2000AJ....120.2131Z} PMs.} (\textit{e}){: Derived using PPMXL catalog \citep{2010AJ....139.2440R}.} (\textit{f}){: Obtained from \citet{2000AA...360..472V} catalog after correcting the PMs for the perspective and solid-body rotations. The errors are the original uncertainties of \citet{2000AA...360..472V} because representation of the sources of systematics in their PMs.} (\textit{g}){: Obtained from \citet{2000AA...360..472V} catalog after removing perspective-, solid-body- and cluster-rotation contributions. The errors are the same of vL00 (@F1).} (\textit{h}){: PM estimate obtained considering all likely cluster members in the catalog, not only those at the position of our field F1.} (\textit{i}){: Obtained after the correction of the perspective and cluster rotations.}}
\end{table*}

\begin{figure*}[!]
  \centering
  \includegraphics[trim=20 150 30 90,clip=true,width=1.05\columnwidth]{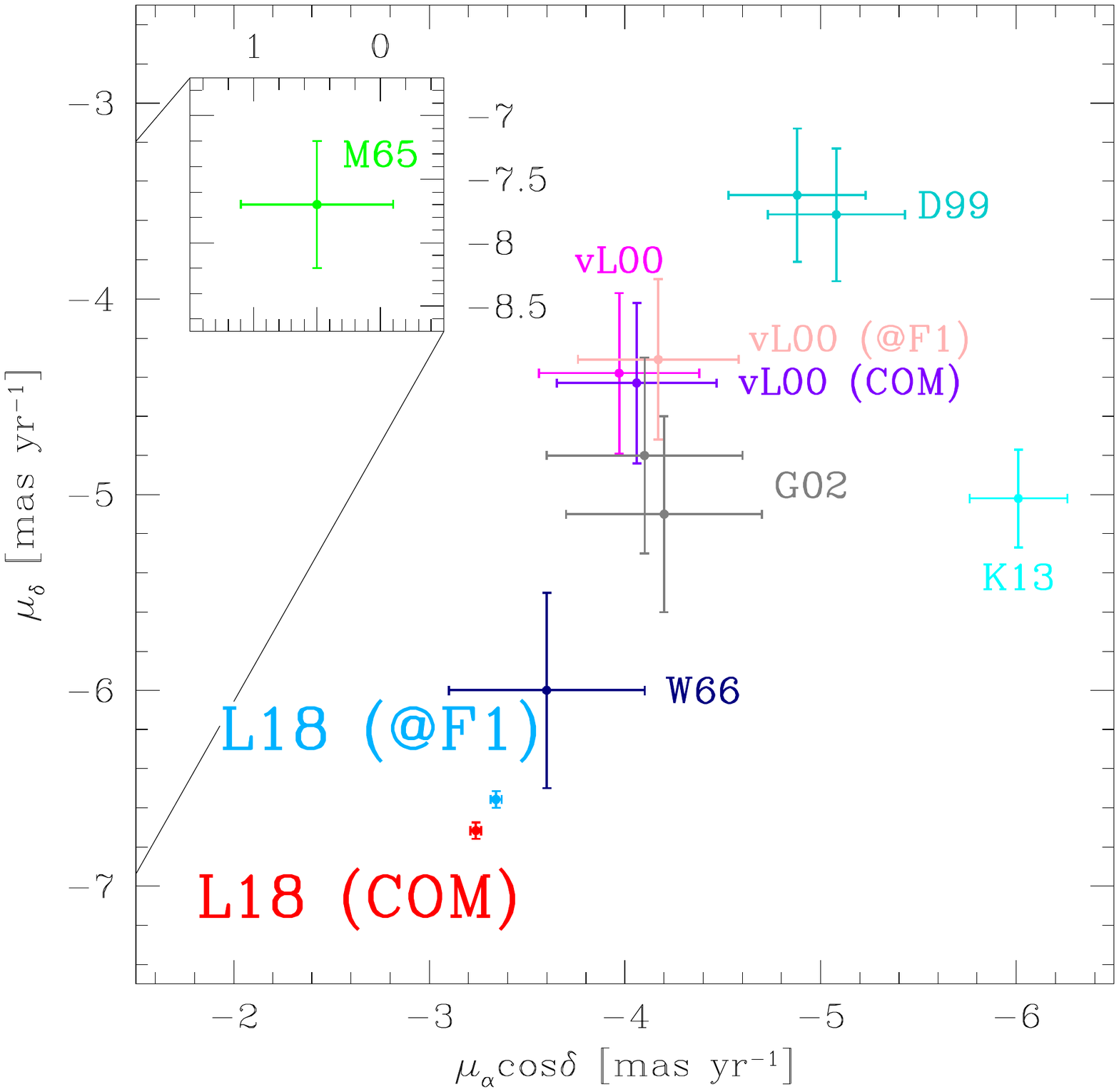}
  \includegraphics[trim=20 150 30 90,clip=true,width=1.05\columnwidth]{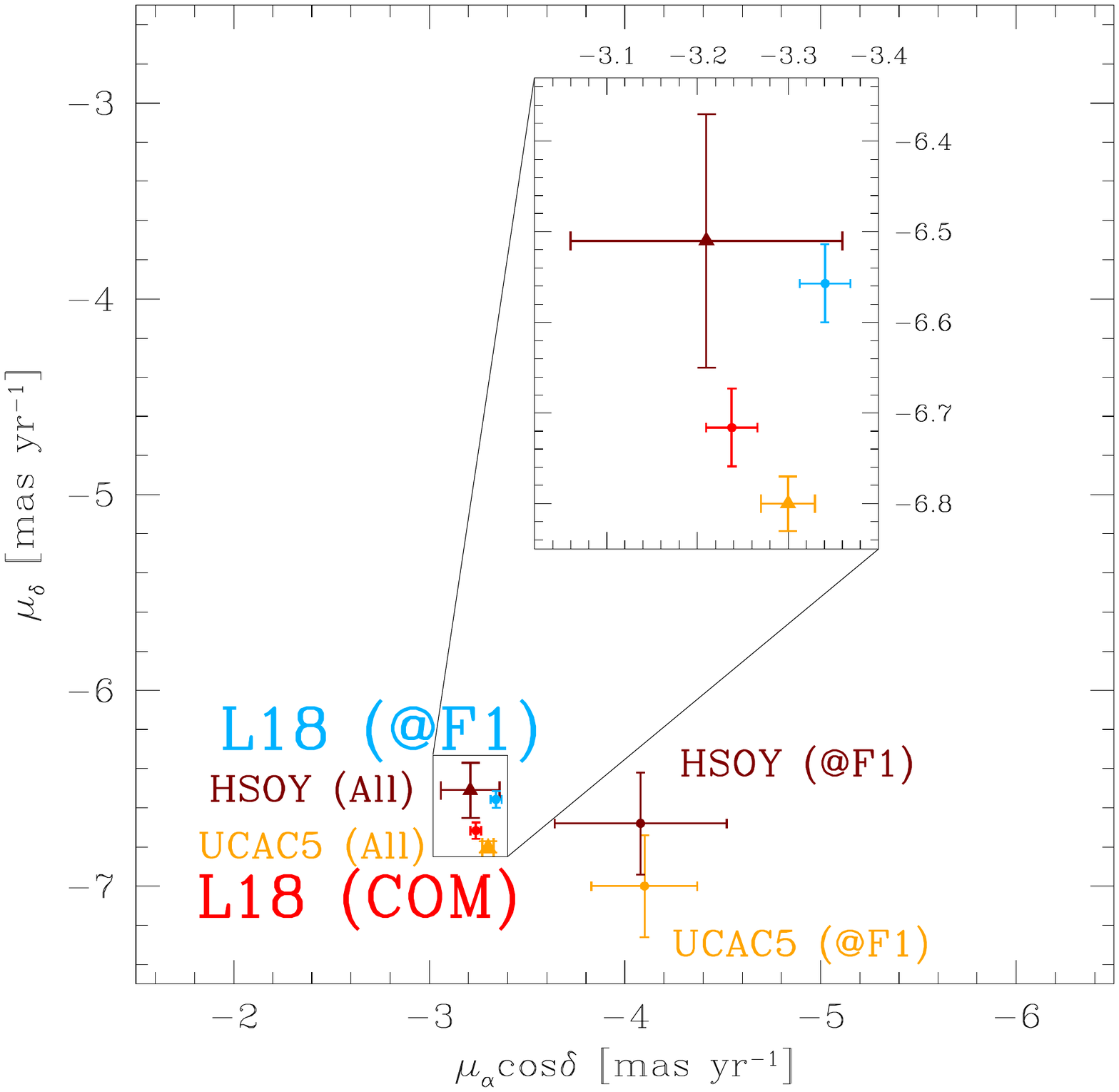}
  \caption{\textit{Left}: comparison between the absolute PM of \ocen
    obtained with \hsts data and the literature. The azure and red
    points represent the absolute PM at the distance of our field (L18
    @F1) and of the COM (L18 COM, obtained with the corrections
    described in Sect.~\ref{rot}), respectively. Using the IDs listed
    in Table~\ref{tab1}, we plot the values of
    \citetalias{1966ROAn....2....1W} in dark blue,
    \citetalias{1999AJ....117..277D} in turquoise,
    \citetalias{2000AA...360..472V} in magenta,
    \citetalias{2002ASPC..265..399G} in gray and
    \citetalias{2013AA...558A..53K} in cyan. The value of
    \citetalias{1965RGOB..100...81M} is shown as a light-green point
    in the inset. The value of \ocen COM absolute PM obtained from the
    corrected \citetalias{2000AA...360..472V} PMs (see text for
    detail) is shown in purple. The value obtained using stars in
    \citetalias{2000AA...360..472V} catalog (with PMs corrected only
    for perspective and solid-body rotations) at the same position of
    our F1 field (in pink) is plot as reference. \textit{Right}:
    absolute-PM of \ocen at the position of our field F1 on the same
    scale obtained by using UCAC5 (in orange) and HSOY (in brown)
    catalogs. The error bars are the standard errors to the mean. The
    orange and brown triangles represent the absolute PM of \ocen
    computed using the PMs of all likely cluster members (not only
    those in the field F1) in UCAC5 and HSOY catalogs,
    respectively. In the inset, we zoom-in around the location of our
    absolute-PM estimates.}\label{fig3}
\end{figure*}

\section{Comparison with the literature}\label{lit}

The current estimates of the absolute PMs of \ocen available in the
literature are collected in Table~\ref{tab1} and are shown in the
left-hand panel of Fig.~\ref{fig3}. With the interesting exception of
the value of \citet{2002ASPC..265..399G} obtained by linking the PM
catalog of \citet{1965RGOB..100...81M} and \citet{1966ROAn....2....1W}
to the Tycho 2 catalog (dark-blue point with label W66), the
disagreement between our result and the literature values is
clear. For example, there is a difference of more than 2 mas yr
$^{-1}$ in modulus and about 30$^\circ$ and 15$^\circ$ in orientation,
resulting in 9$\sigma$ and 5$\sigma$ discrepancies, between our value
and those of \citetalias{1999AJ....117..277D} and
\citetalias{2000AA...360..472V}, respectively.

To understand whether our methodology or the analyzed field can be at
the basis of these discrepancies, as test we downloaded the
publicly-available \citetalias{2000AA...360..472V} catalog and
computed the absolute PM of \ocen using PMs therein. We limited our
analysis to 45 cluster members (membership probability greater than
85\% in the \citetalias{2000AA...360..472V} catalog) at the position
of our field F1.

\citetalias{2006AA...445..513V} pointed out that the PMs of
\citetalias{2000AA...360..472V} are affected by perspective and
solid-body (due to a misalignment between the photographic plates of
different epochs) rotations. Therefore, we corrected the perspective
component as described in Sect.~\ref{rot}; and the solid-body part as
described in \citetalias{2006AA...445..513V} (Eq.~7), using the
average position of the 45 stars as reference, and transforming the
corresponding PM correction from the \ocen axisymmetric reference
frame to the equatorial reference frame. By applying these two
corrections, we computed the absolute PM at the position of our field
F1 (pink point in the left-hand panel of Fig.~\ref{fig3}). Finally, we
removed the effect of the cluster rotation as described in
Sect.~\ref{rot} and obtained the absolute PM of \ocen COM. The result
is shown in the left-hand panel of Fig.~\ref{fig3} as a purple
point. Again, the difference with our absolute PM estimates is about
5$\sigma$.

We also measured the absolute PM of \ocen using the absolute PMs
published in the UCAC5 and ``Hot Stuff for One Year''
\citep[HSOY,][]{2017AA...600L...4A} catalogs. We chose these two
catalogs because both of them employ the Gaia DR1 catalog to link
their astrometry to an absolute reference-frame system. The
absolute-PM measurements were obtained by using 45 cluster members,
selected on the basis of their location in the CMD and in the VPD, at
the same location of the stars analyzed in our paper. In the
right-hand panel of Fig.~\ref{fig3} we show the absolute PM of \ocen
at the distance of our field F1 (i.e., without considering the effect
of the perspective and cluster rotation) for UCAC5 (orange point) and
HSOY (brown point) catalogs. Both UCAC5 and HSOY values are closer to
our measurements than to those in the literature. For completeness, we
also computed the PM of \ocen using all likely cluster members in the
UCAC5 and HSOY catalogs, not only those at the position of our field
F1. By considering a more populated sample in the computation (2394
for UCAC5 and 339 for HSOY), the agreement with our estimate becomes
tighter (orange and brown triangles for UCAC5 and HSOY, respectively,
in the right-hand panel of Fig.~\ref{fig3}). Note that the errors for
UCAC5 value are smaller than those for our estimate because of the
larger number of considered objects. This external check further
strengthens our result.

\begin{table*}
  \begin{center}
    \caption{Parameters employed for the Galactic Potential.} \label{tab2}
    \begin{tabular}{lcr}
      \tableline
      \tableline
      \textbf{Parameter} & \textbf{Value} & \textbf{References} \\
      \tableline
      \tableline
      $R_0$            & $8.30 \pm 0.23$ kpc             & (\textit{a}) \\
      ${\Theta}_0$     & $239 \pm 7$ km s$^{-1}$                  & (\textit{a}) \\
      $(U,V,W)_{\odot}$ & $(-11.10 \pm 1.20,12.24 \pm 2.10,7.25 \pm 0.60)$ km s$^{-1}$ & (\textit{a,b}) \\
      \tableline
      \multicolumn{3}{c}{\textit{Galactic Bar}} \\
      Mass                           & $1.1 \times 10^{10}$ M$_{\odot}$   & (\textit{c,d}) \\
      Present position of major axis & 20$^{\circ}$                      & (\textit{e}) \\
      $\Omega_{\rm bar}$               & 40, 45, 50 km s$^{-1}$ kpc$^{-1}$ & (\textit{f,g}) \\
      Cut-off radius                 & 3.28 kpc                        & (\textit{d,h,i})\\ 
      \tableline
      \multicolumn{3}{c}{\textit{Spiral Arms}} \\
      $M_{\rm arms}/M_{\rm disk}$ &  0.05                    & (\textit{l}) \\
      Scale length            & 5.0 kpc                 & (\textit{m}) \\
      Pitch angle             & 15.5$^{\circ}$            & (\textit{n}) \\
      Angular velocity        & 25 km s$^{-1}$ kpc$^{-1}$ & (\textit{m,o}) \\
      \tableline
      \tableline
    \end{tabular}
  \end{center}
  \tablecomments{References. (\textit{a}){: \citet{2011AN....332..461B}.} (\textit{b}){: \citet{2010MNRAS.403.1829S}.} (\textit{c}){: \citet{2012AA...538A.106R}.} (\textit{d}){: \citet{2017arXiv170805742F}.} (\textit{e}){: \citet{2002ASPC..273...73G}.} (\textit{f}){: \citet{2003MNRAS.340..949B}.} (\textit{g}){: \citet{2015MNRAS.454.1818S}.} (\textit{h}){: \citet{2017arXiv171007433F}.} (\textit{i}){: \citet{2015MNRAS.448..713P}.} (\textit{l}){: \citet{2012AJ....143...73P}.} (\textit{m}){: \citet{2017MNRAS.468.3615M}.} (\textit{n}){: \citet{2000AA...358L..13D}.} (\textit{o}){: \citet{2011MSAIS..18..185G}.} }

\vskip 20pt

  \begin{center}
    \caption{\ocen parameters adopted in the orbit computation.} \label{tabkin}
    \begin{tabular}{lcr}
      \tableline
      \tableline
      \textbf{Parameter} & \textbf{Value} & \textbf{Reference} \\
      \tableline
      \tableline
      $(\alpha,\delta)_{\rm J2000}$ & ($13^{\rm h}26^{\rm m}47^{\rm s}\!\!.24$,$-47^\circ28^\prime46^{\prime\prime}\!\!.45$) & (\textit{a}) \\
      Radial Velocity & $232.2 \pm 0.7$ km s$^{-1}$ & (\textit{b,c}) \\
      Distance & $5.20 \pm 0.25$ kpc & (\textit{d}) \\
      $X$ & $-5.1$ kpc \\
      $Y$ & $-3.9$ kpc \\
      $Z$ & 1.3 kpc \\
      \tableline
      \multicolumn{3}{c}{\citet{1999AJ....117..277D}} \\
      $\mu_\alpha \cos\delta$ & $(-5.08 \pm 0.35)$ mas yr$^{-1}$ \\
      $\mu_\delta$ & $(-3.57 \pm 0.34)$ mas yr$^{-1}$ \\
      $v_X$ &   58.0  km s$^{-1}$ \\
      $v_Y$ & $-22.8$ km s$^{-1}$ \\
      $v_Z$ &    0.2  km s$^{-1}$ \\
      \tableline
      \multicolumn{3}{c}{Our paper (L18 @F1) (\textit{$^\ast$})} \\
      $\mu_\alpha \cos\delta$ & $(-3.341 \pm 0.028)$ mas yr$^{-1}$ \\
      $\mu_\delta$ & $(-6.557 \pm 0.043)$ mas yr$^{-1}$ \\
      $v_X$ &   95.7  km s$^{-1}$ \\
      $v_Y$ & $-18.4$ km s$^{-1}$ \\
      $v_Z$ & $-76.1$ km s$^{-1}$ \\
      \tableline
      \tableline
    \end{tabular}
  \end{center}
  \tablecomments{(\textit{$^\ast$}): We adopted the F1 PM value instead that of the COM to remove any dependence on the correction for the PM rotation of \ocen. References. (\textit{a}){: \citet{2010PASP..122.1035A}.} (\textit{b}){: \citet{1997AJ....114.1087M}.}  (\textit{c}){: \citet{2006A&A...445..503R}.} (\textit{d}){: \citet{1996AJ....112.1487H}, 2010 edition. We adopted a distance error of $\sim 5$\%, which is a reasonable guess in agreement with the most-recent estimates for \ocen distance by using RR Lyrae stars \citep[e.g.,][]{2015A&A...574A..15F,2017A&A...606C...1N}}}

\end{table*}

\section{The orbit of \ocen}\label{orbit}

In order to analyze the repercussions of the new PM determinations on
the orbit of \ocen, we computed the orbit in two detailed mass models
based on the Milky Way Galaxy.

The first model is based on an escalation of the axisymmetric Galactic
model of \citet{1991RMxAA..22..255A}, for which we converted its bulge
component into a prolate bar with the mass distribution given in
\citet{2004ApJ...609..144P}. We also included the 3D spiral-arm model
considered in \citet{2003ApJ...582..230P} to explore any possible
effect of the spiral arms on \ocen. For this first potential, we
computed the minimum and maximum distances from the Galactic center,
the maximum vertical distance from the Galactic plane, the orbital
eccentricity, and the angular momentum. The eccentricity $e$ is
defined as:
\vskip -8 pt
\begin{equation}
  e = (R_{\rm max}-R_{\rm min})/(R_{\rm max}+R_{\rm min}) \textrm{ ,}
\end{equation}
where $R_{\rm min}$ and $R_{\rm max}$ are successive minimum and
maximum distances from the Galactic rotation axis (i.e., $R$ is the
distance in cylindrical coordinates), respectively.

From a gravitational perspective, the most influential part of the
Galaxy for dynamically hot (eccentric) stellar clusters or dwarf
galaxies is the inner region, where the bar and the maximum mass
concentration of the Galaxy are located. An accurate model of the
central part of the Galaxy is paramount for an accurate determination
of the orbit. For these reasons, we performed a second experiment
using the new galaxy-modeling algorithm called
\texttt{GravPot16}\footnote{\href{https://fernandez-trincado.github.io/GravPot16/}{https://fernandez-trincado.github.io/GravPot16/}}
\citep{2017arXiv170805742F,FT18} that specifically attempts to model
the inner Galactic region. \texttt{GravPot16} is a semi-analytic,
steady-state, 3D gravitational potential of the Milky Way
observationally and dynamically constrained. \texttt{GravPot16} is a
non-axisymmetric potential including only the bar in its
non-axisymmetric components. The potential model is primarily the
superposition of several composite stellar components where the
density profiles in cylindrical coordinates, $\rho_{i} (R,z)$, are the
same as those proposed in
\citet{2003AA...409..523R,2012AA...538A.106R,2014AA...569A..13R},
i.e., a boxy/peanut bulge, a Hernquist stellar halo, seven stellar
Einasto thin disks with spherical symmetry in the inner regions, two
stellar sech$^{2}$ thick disks, a gaseous exponential disk, and a
spherical structure associated with the dark matter halo. A new
formulation for the global potential of this Milky Way density model
will be described in detail in a forthcoming paper \citep{FT18}. For
the second potential, we computed the orbits in a Galactic
non-inertial reference frame in which the bar is at rest, and we
specifically examined the orbital Jacobi constant per unit mass,
$E_{\rm J}$, which is conserved in this reference frame. The orbital
energy $E_{\rm J}$ per unit of mass is defined as:
\begin{equation}
  E_{\rm J} = \frac{1}{2}\vec{v}^2 + \Phi_{\rm axi} + \Phi_{\rm non-axi} - \frac{1}{2}| \vec{\Omega}_{\rm bar}\times \vec{R} |^2 \textrm{ ,}
\end{equation}
with $\Phi_{\rm axi}$ and $\Phi_{\rm non-axi}$ the gravitational
potential of the axisymmentric and non-axisymmentric (in this case the
bar) Galactic potential, respectively; $\Omega_{\rm bar}$ the angular
velocity gradient, and $R$ the vector $\vec{r}$ of the
particles/stars.

For the potential of the bar, we assumed (i) a total mass of
$1.1\times10^{10}$ M$_{\odot}$, (ii) an angle of 20$^{\circ}$ for the
present-day orientation of the major axis of the bar, (iii) an angular
velocity gradient of the bar $\Omega_{\rm bar}$ of 40, 45, and 50 km
s$^{-1}$ kpc$^{-1}$, and (iv) a cut-off radius for the bar of $R_{\rm
  cut} = 3.28$ kpc. The bar potential model was computed using a new
mathematical technique that considers ellipsoidal shells with similar
linear density \citep{FT18}. Table~\ref{tab2} summarizes the
parameters employed for the Galactic potential, the position and
velocity of the Sun in the Local Standard of Rest, and the properties
of the bar and of the spiral arms.

\begin{figure*}
  \centering
  \includegraphics[trim=20 140 00 90,clip=true,scale=0.5,keepaspectratio]{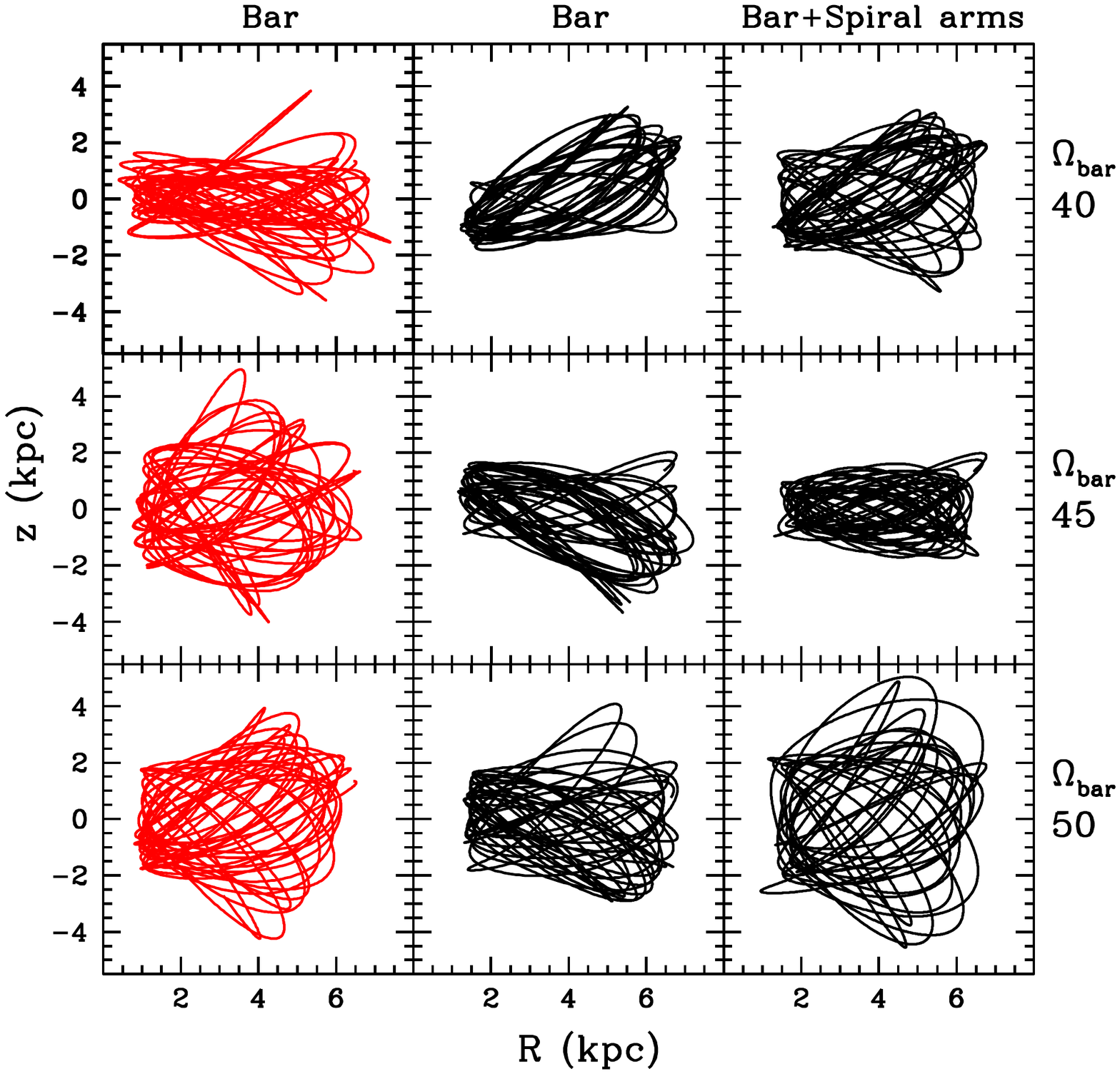}
  \caption{Meridional orbits of \ocen for the last 2 Gyr for our first
    Galactic potential, with an angular velocity gradient of the
    Galactic bar of 40 (top row), 45 (middle row), 50 (bottom row) km
    s$^{-1}$ kpc$^{-1}$. The orbits in the first column (red lines)
    employed the PM value of \citetalias{1999AJ....117..277D}, while
    the orbits in the second and third columns (black lines) adopted
    those computed in this paper. In the third column, the orbits were
    computed by adding the spiral arms with an angular velocity of 25
    km s$^{-1}$ kpc$^{-1}$.}\label{Model1}

  \centering
  \includegraphics[trim=20 140 00 90,clip=true,scale=0.5,keepaspectratio]{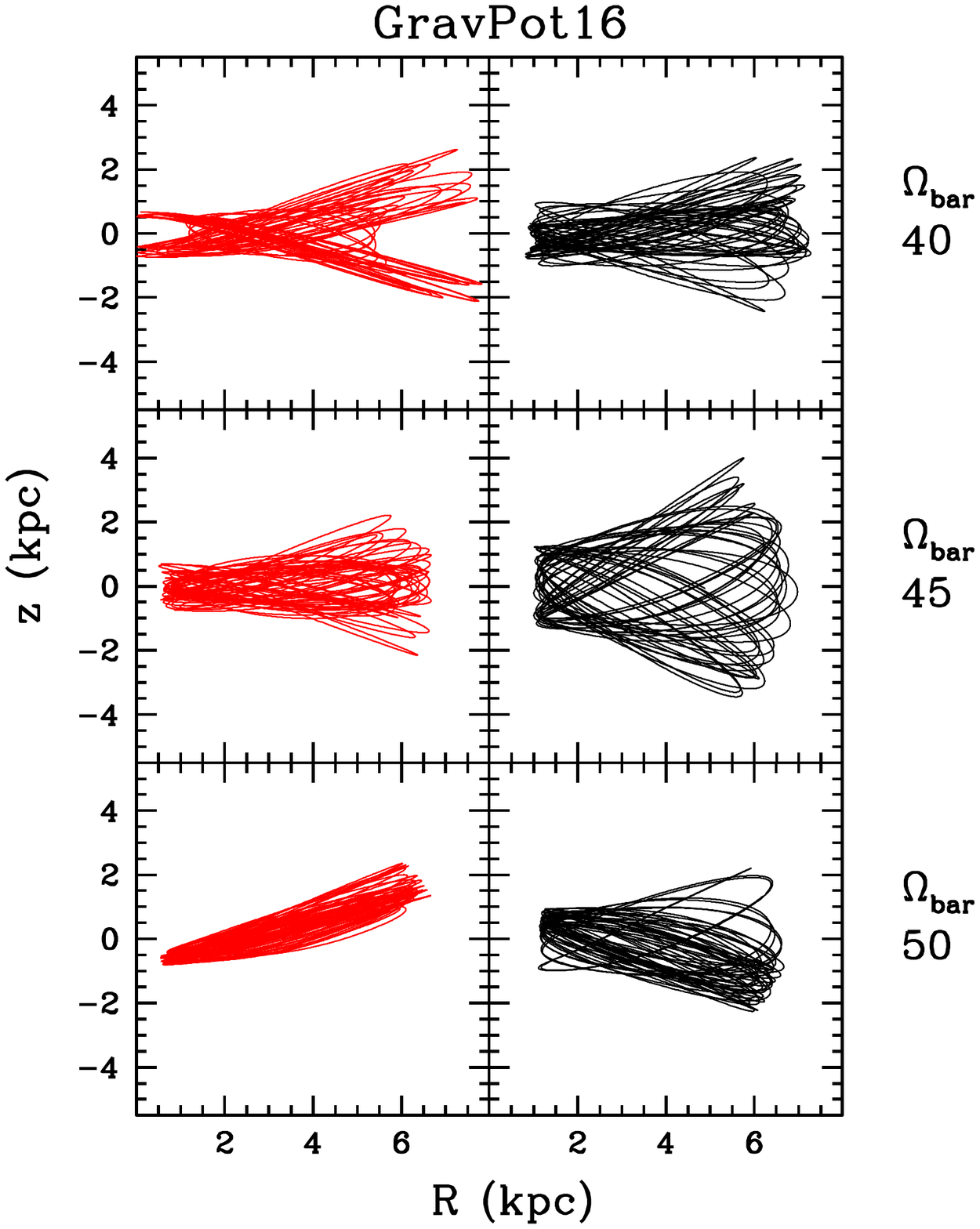}
  \caption{Same as Fig.~\ref{Model1} but with orbits calculated
    assuming the second Galactic potential..}\label{Model2}
\end{figure*}

Hereafter, we discuss the results achieved by employing the
absolute-PM value measured for the stars in field F1. We also ran the
corresponding calculations for the estimated COM PM and found a
negligible difference (e.g., comparable perigalacticons,
apogalacticons, region covered by the orbit) as expected given the
small PM difference between the two estimates. The COM is physically
the correct quantity to use, but by using the value for F1 we remove
any dependence on the assumed, and possibly uncertain, correction for
the PM rotation of the cluster (and note that the perspective and
rotation corrections for \ocen are smaller than the uncertainty in the
azimuthal velocity of the Sun). The other \ocen parameters are listed
in Table~\ref{tabkin}. Finally, we also calculated, for comparison,
the \ocen orbit by assuming the PM value of
\citetalias{1999AJ....117..277D}.

Figures~\ref{Model1} and \ref{Model2} present the orbits of \ocen
obtained by considering the first and the second Galactic potential,
respectively. For each value of the angular velocity gradient of the
Galactic bar (different rows) we investigated, we show the meridional
orbit of \ocen obtained using the absolute-PM value of
\citetalias{1999AJ....117..277D} (red lines in the left columns) and
that computed with \hsts data in our paper (black lines in the other
columns). Furthermore, in the rightmost column of Fig.~\ref{Model1},
we display the orbit calculated by including the spiral arms in the
computation.

Looking at Fig.~\ref{Model1}, some differences can be discerned
between the different orbits. The radial excursions of the orbit of
\ocen seem to increase slightly with the new PM value, while the
vertical excursions remain similar regardless of the PM adopted. The
main difference between the new and the literature PM values is the
larger (about half a kpc) perigalactic distance obtained with the PM
computed in our paper. This more distant approach to the innermost
part of the Galaxy likely means a larger survival expectancy for the
cluster.

Examining the impact of the spiral arms, contrary to intuition we find
that the spiral arms can influence the orbit, although
slightly. However, the differences with this and the bar-only
potential model is statistical and does not change the fundamental
properties of the orbit (i.e., perigalacticons and apogalacticons).

Finally, by comparing Figs.~\ref{Model1} and \ref{Model2}, we found
that both models seem to reproduce a similar behavior in the radial
and vertical excursions, and no important differences due to the
details of the modeling are detectable. This means that probably the
most important features on models that influence the orbital behavior
of this particular cluster are the density law and the angular
velocity of the bar that are basically the same in both models.

In Tables~\ref{tab3} and \ref{tab4} we present average orbital
parameters for \ocen for the various Galactic potentials obtained by
using a simple Monte Carlo procedure to vary the initial
conditions. 500 orbits for the first potential and 50\,000 orbits for
the second potential are computed under 1$\sigma$ variation of the
absolute PMs, distance, radial velocity, and Solar motion. The errors
provided for these parameters were assumed to follow a Gaussian
distribution. The parameters listed in the Tables are the minimum and
maximum distances to the Galactic center, the maximum distance from
the Galactic plane, and the orbital eccentricity. In Table~\ref{tab3}
we also provide the angular momentum, while in Table~\ref{tab4} we
give the orbital Jacobi energy.

\section{Conclusions}

As part of the \hsts large program GO-14118+14662 on \ocen, in this
third paper of the series we presented a revised estimate of the
absolute PM of this cluster. Thanks to the exquisite \hsts-based
astrometry and high-precision PMs computed in
\citetalias{2018arXiv180101504B}, we are able to use background
galaxies to link our relative astrometry to an absolute frame. Our
measured absolute PM differs significantly from the values available
in the literature in magnitude, direction and precision. A comparison
with recent catalogs yields instead an excellent agreement.

We also calculated the orbit of \ocen with two different models of the
potential for the Milky Way Galaxy. The primary difference between the
orbits with our new PM and those based on prior estimates is a larger
(by about half a kpc) perigalactic distance obtained with our new
value. This may imply a larger survival expectancy for the cluster in
the inner Galactic environment.

We emphasize that the absolute PM presented here is both an
independent (with the exception of the registration of the
master-frame absolute scale and orientation with the Gaia DR1 catalog)
and direct measurement based on the use of a reference sample of
background galaxies. As such, these results presented here provide an
external check for future astrometric all-sky catalogs, in particular
the upcoming Gaia DR2.

\section*{Acknowledgments}

We thank the anonymous referee for the useful comments and
suggestions. ML, AB, DA, AJB and JMR acknowledge support from STScI
grants GO 14118 and 14662. JGF-T gratefully acknowledges the Chilean
BASAL Centro de Excelencia en Astrof\'isica y Tecnolog\'ias Afines
(CATA) grant PFB- 06/2007. Funding for the GravPot16 software has been
provided by the Centre national d'\'etudes spatiale (CNES) through
grant 0101973 and UTINAM Institute of the Universit\'e de
Franche-Comte, supported by the Region de Franche-Comte and Institut
des Sciences de l'Univers (INSU). APM acknowledges support by the
European Research Council through the ERC-StG 2016 project 716082
'GALFOR'. Based on observations with the NASA/ESA \hstl, obtained at
the Space Telescope Science Institute, which is operated by AURA,
Inc., under NASA contract NAS 5-26555. This work has made use of data
from the European Space Agency (ESA) mission {\it Gaia}
(\url{https://www.cosmos.esa.int/gaia}), processed by the {\it Gaia}
Data Processing and Analysis Consortium (DPAC,
\url{https://www.cosmos.esa.int/web/gaia/dpac/consortium}). Funding
for the DPAC has been provided by national institutions, in particular
the institutions participating in the {\it Gaia} Multilateral
Agreement.

\begin{table*}
  \begin{center}
    \caption{Monte-Carlo, average orbital parameters (and
      corresponding $1\sigma$ uncertainties) of \ocen for our first
      Galactic Potential, using \citetalias{1999AJ....117..277D} and
      our absolute PM (@F1) values. For each angular velocity gradient
      of the Galactic bar, the first and the second rows represent the
      potential with only the bar and with bar $+$ spiral arms,
      respectively. From left to right, we provide: the angular
      velocity of the Galactic bar ($\Omega_{\rm bar}$), the minimum
      ($\langle r_{\rm min} \rangle$) and maximum ($\langle r_{\rm
        max} \rangle$) distances to the Galactic center, the maximum
      distance from the Galactic plane ($\langle |z|_{\rm max}
      \rangle$, the orbital eccentricity ($\langle e \rangle$), and
      the angular momentum ($\langle h \rangle$).} \label{tab3}
    \begin{tabular}{cccccc}
      \tableline
      \tableline
      \textbf{$\bf{\Omega}_{\rm bar}$} & \textbf{$\bf \langle$$\bf r_{\rm min}$$\bf \rangle$} & \textbf{$\bf \langle$$\bf r_{\rm max}$$\bf \rangle$} & \textbf{$\bf \langle$$\bf |z|_{\rm max}$$\bf \rangle$} & \textbf{$\bf \langle$$\bf e$$\bf \rangle$} & \textbf{$\bf \langle$$\bf h$$\bf \rangle$} \\
      \textbf{[km s$^{-1}$ kpc$^{-1}$]} & \textbf{[kpc]} & \textbf{[kpc]} & \textbf{[kpc]} & & \textbf{[km s$^{-1}$ kpc]} \\
      \tableline
      \tableline
      \multicolumn{6}{c}{\textbf{\citet{1999AJ....117..277D}}} \\
      40 & $0.78 \pm 0.16$ & $7.39 \pm 0.35$ & $3.86 \pm 0.28$ & $0.76 \pm 0.05$ & $-319 \pm 54$ \\
         & $0.83 \pm 0.23$ & $7.17 \pm 0.45$ & $3.78 \pm 0.40$ & $0.71 \pm 0.07$ & $-351 \pm 64$ \\
      ~\\
      45 & $0.83 \pm 0.12$ & $7.12 \pm 0.32$ & $3.81 \pm 0.69$ & $0.72 \pm 0.05$ & $-341 \pm 41$ \\
         & $0.84 \pm 0.18$ & $7.08 \pm 0.26$ & $3.70 \pm 0.63$ & $0.72 \pm 0.05$ & $-346 \pm 49$ \\
      ~\\
      50 & $0.91 \pm 0.13$ & $6.86 \pm 0.29$ & $3.67 \pm 0.52$ & $0.69 \pm 0.04$ & $-373 \pm 39$ \\
         & $0.92 \pm 0.15$ & $6.83 \pm 0.24$ & $3.63 \pm 0.60$ & $0.69 \pm 0.04$ & $-368 \pm 41$ \\
      \tableline
      \multicolumn{6}{c}{\textbf{Our paper (L18 @F1)}} \\
      40 & $1.39 \pm 0.14$ & $7.21 \pm 0.20$ & $3.52 \pm 0.59$ & $0.61 \pm 0.03$ & $-513 \pm 42$ \\
         & $1.45 \pm 0.19$ & $7.13 \pm 0.16$ & $3.39 \pm 0.66$ & $0.57 \pm 0.04$ & $-538 \pm 56$ \\
      ~\\
      45 & $1.23 \pm 0.16$ & $7.53 \pm 0.58$ & $3.65 \pm 0.78$ & $0.64 \pm 0.05$ & $-495 \pm 36$ \\
         & $1.48 \pm 0.14$ & $7.06 \pm 0.13$ & $2.94 \pm 0.82$ & $0.57 \pm 0.04$ & $-565 \pm 50$ \\
      ~\\
      50 & $1.32 \pm 0.16$ & $7.40 \pm 0.31$ & $4.26 \pm 0.55$ & $0.62 \pm 0.04$ & $-475 \pm 49$ \\
         & $1.36 \pm 0.20$ & $7.30 \pm 0.36$ & $4.03 \pm 0.76$ & $0.59 \pm 0.04$ & $-494 \pm 58$ \\
      \tableline
      \tableline
    \end{tabular}

\vskip 20pt

    \caption{Same as for Table~\ref{tab3} but using the second
      Galactic Potential. The orbital parameters here provided are
      referred to the non-inertial reference frame where the bar is at
      rest. For this reason, we provide the orbital Jacobi energy
      $E_{\rm J}$ instead of the angular momentum. The superscript and
      subscript values represent the upper (84$^{\rm th}$ percentile)
      and lower (16$^{\rm th}$ percentile) limits,
      respectively.} \label{tab4}
    \begin{tabular}{cccccc}
      \tableline
      \tableline
      \textbf{$\bf{\Omega}_{\rm bar}$} & \textbf{$\bf \langle$$\bf r_{\rm min}$$\bf \rangle$} & \textbf{$\bf \langle$$\bf r_{\rm max}$$\bf \rangle$} & \textbf{$\bf \langle$$\bf |z|_{\rm max}$$\bf \rangle$} & \textbf{$\bf \langle$$\bf e$$\bf \rangle$} & \textbf{$\bf \langle$$\bf E_{\rm J}$$\bf \rangle$} \\
      \textbf{[km s$^{-1}$ kpc$^{-1}$]} & \textbf{[kpc]} & \textbf{[kpc]} & \textbf{[kpc]} & & \textbf{[10$^2$ km$^2$ s$^{-2}$]} \\
      \tableline
      \tableline
      \multicolumn{6}{c}{\textbf{\citet{1999AJ....117..277D}}} \\
      40 & $0.10^{0.20}_{0.02}$ & $7.80^{8.13}_{7.48}$ & $2.61^{2.77}_{2.44}$ & $0.98^{0.99}_{0.95}$ & $-1949.62^{-1939.96}_{-1959.09}$ \\
      ~\\
      45 & $0.48^{0.59}_{0.38}$ & $6.80^{6.91}_{6.68}$ & $2.43^{2.61}_{2.29}$ & $0.87^{0.90}_{0.84}$ & $-1935.27^{-1924.63}_{-1945.71}$ \\
      ~\\
      50 & $0.55^{0.63}_{0.47}$ & $6.64^{6.73}_{6.55}$ & $2.38^{2.48}_{2.27}$ & $0.85^{0.87}_{0.83}$ & $-1920.95^{-1909.32}_{-1932.38}$ \\
      \tableline
      \multicolumn{6}{c}{\textbf{Our paper (L18 @F1)}} \\
      40 & $0.90^{1.07}_{0.79}$ & $7.12^{7.30}_{6.87}$ & $2.81^{3.21}_{2.36}$ & $0.77^{0.80}_{0.73}$ & $-1842.60^{-1825.52}_{-1859.50}$ \\
      ~\\
      45 & $0.95^{1.07}_{0.83}$ & $7.04^{7.20}_{6.91}$ & $3.05^{3.59}_{2.55}$ & $0.76^{0.79}_{0.73}$ & $-1822.04^{-1803.51}_{-1840.40}$ \\
      ~\\
      50 & $1.04^{1.17}_{0.88}$ & $6.83^{7.08}_{6.60}$ & $2.84^{3.35}_{2.35}$ & $0.74^{0.78}_{0.70}$ & $-1801.47^{-1781.43}_{-1821.23}$ \\
      \tableline
    \end{tabular}
  \end{center}
\end{table*}

\bibliographystyle{aasjournal}



\appendix

\section{Galaxy finding charts}\label{A1}

\begin{figure*}
  \centering
  \includegraphics[trim=320 0 320 0,clip=true,scale=5.0,keepaspectratio]{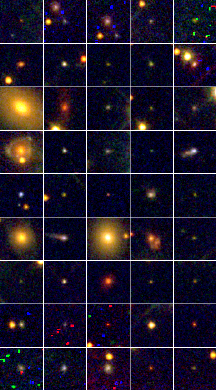}
  \caption{Trichromatic stacked images of the galaxies adopted in our
    analysis.}\label{fig1a}
\end{figure*}

In Fig.~\ref{fig1a} we show the trichromatic stacked images of the 45
galaxies used to compute the absolute PM of \ocen. Each source is
centered in a $51 \times 51$ WFC3/UVIS pixel$^2$ ($\sim 2 \times 2$
arcsec$^2$) stamp. We used F814W, F606W and F336W filters for red,
green and blue channels, respectively.

\end{document}